\documentclass[10pt,twocolumn,letterpaper]{article}

\usepackage{iccv}
\usepackage{times}
\usepackage{epsfig}
\usepackage{graphicx}
\usepackage{amsmath}
\usepackage{amssymb}

\usepackage{booktabs} 
\usepackage{multirow} 
\usepackage{caption}
\usepackage{subcaption}

\usepackage[breaklinks=true,bookmarks=false]{hyperref}
\iccvfinalcopy 

\def\iccvPaperID{****} 

\ificcvfinal\fi

\begin{document}

\title{Improving Automatic Endoscopic Stone Recognition Using a Multi-view Fusion Approach Enhanced with Two-Step Transfer Learning}
\author{Francisco Lopez-Tiro$^{1,2}$
\and
Elias Villalvazo-Avila$^{1}$
\and
Juan Pablo Betancur-Rengifo$^{1}$
\and
Ivan Reyes-Amezcua$^{3}$
\and
Jacques Hubert$^{4}$
\and
Gilberto Ochoa-Ruiz$^{1}$
\and
Christian Daul$^{2}$
\and \\ 
$^{1}$Tecnologico de Monterrey, School of Engineering and Sciences, Mexico\\
$^{2}$CRAN UMR 7039, Université de Lorraine and CNRS, Nancy, France\\
$^{3}$CINVESTAV, Guadalajara, Mexico\\
$^{4}$CHRU Nancy, Service d’urologie de Brabois, Nancy, France\\ \\
}

\maketitle
\ificcvfinal\thispagestyle{empty}\fi

\begin{abstract}
This contribution presents a deep-learning method for extracting and fusing image information acquired from different viewpoints, with the aim to produce more discriminant object features for the identification of the type of kidney stones seen in endoscopic images. 
The approach was specifically designed to mimic the morpho-constitutional analysis to visually classify kidney stones by jointly using surface and section images of kidney stone fragments. 
The model was further improved with a two-step transfer learning approach and by attention blocks to refine the learned feature maps. Deep feature fusion strategies improved the results of single view extraction backbone models by more than 6\% in terms of accuracy of the kidney stones classification.
\end{abstract}

\section{Introduction}
\label{sec:intro}

The formation of kidney stones that cannot freely pass through the urinary tract is a major public health issue \cite{cloutier2015kidney, kasidas2004renal, hall2009nephrolithiasis}.
In industrialized countries, it has been reported that at least 10\% of the population suffers from a kidney stone episode once in their lifetime.
In the United States alone, the risk of relapse of the same type of kidney stone has increased by up to 40\% \cite{scales2012prevalence, viljoen2019renal}.  
The formation of kidney stones is caused by different factors such as diet, low fluid intake, and a sedentary lifestyle  \cite{silva2010chemical, daudon2012stone}. However, there are other unavoidable factors such as age, genetic inheritance, and chronic diseases that increase the risk of forming kidney stones \cite{friedlander2015diet}.
Therefore, methods for identifying the different types of kidney stones are crucial for the prescription of appropriate treatments and to reduce the risk of relapses \cite{kartha2013impact, friedlander2015diet}. 
In order to carry out this identification in clinical practice, different procedures have been developed, such as the Morpho-Constitutional Analysis (MCA), and Endoscopic Stone Recognition (ESR) 
\cite{daudon2004clinical, estrade2017pourquoi}.

MCA is commonly accepted as the standard procedure for determining the different types of kidney stones (up to 21 different types and sub-types including pure and mixed compositions are recognized during the MCA) \cite{corrales2021classification}. MCA consists of a double laboratory analysis of kidney stone fragments extracted from the urinary tract during an ureteroscopy \cite{daudon2018recurrence}. 

First, a biologist performs a visual inspection of the kidney stone which is observed with a magnifying glass. This inspection aims to describe kidney stones in terms of colors, textures, and morphology \cite{corrales2021classification}. This visual analysis is done both for the surface view (the external part of the kidney stone fragment), and for a cross-section of the kidney stone fragment (the internal stone part may consist of several layers surrounding a nucleus). Then, the kidney stones are ground up and the resulting powder is used to perform a biochemical analysis using a Fourier Transform Infrared Spectroscopy (FTIR) \cite{khan2018fourier}. The FTIR provides a detailed description of the chemical composition of the kidney stone. 
Finally, the MCA analysis returns the type of kidney stone through a detailed report of the biochemical and morphological characteristics of both views of the kidney stone. 
However, MCA has some major drawbacks: the results are often available only after several weeks, and it is difficult to have a specialized team in each hospital to perform MCA. 

Therefore, urologists have proposed, as a possible alternative, the Endoscopic Stone Recognition (ESR) procedure in which the most common kidney stones are visually identified on the video displayed on a screen during the ureteroscopy itself \cite{estrade2022towards}. 
However, this visual analysis of the surface and section views requires a great deal of expertise due to the high similarities between classes.  Only a limited number of specialists have this expertise. In addition, this technique is more operator dependent and subjective than MCA.
Therefore, new approaches based on deep-learning (DL) methods have been proposed to automate and speed-up the kidney stone identification. Such automated recognition can potentially assists urologists for a real-time decision-making during an ureteroscopy.

\begin{figure*} [] 
    \centering
    \subfloat[Dataset A: CCD-camera images (ex-vivo) ]{\label{fig:dataseta}
    \includegraphics[width=0.45\textwidth]{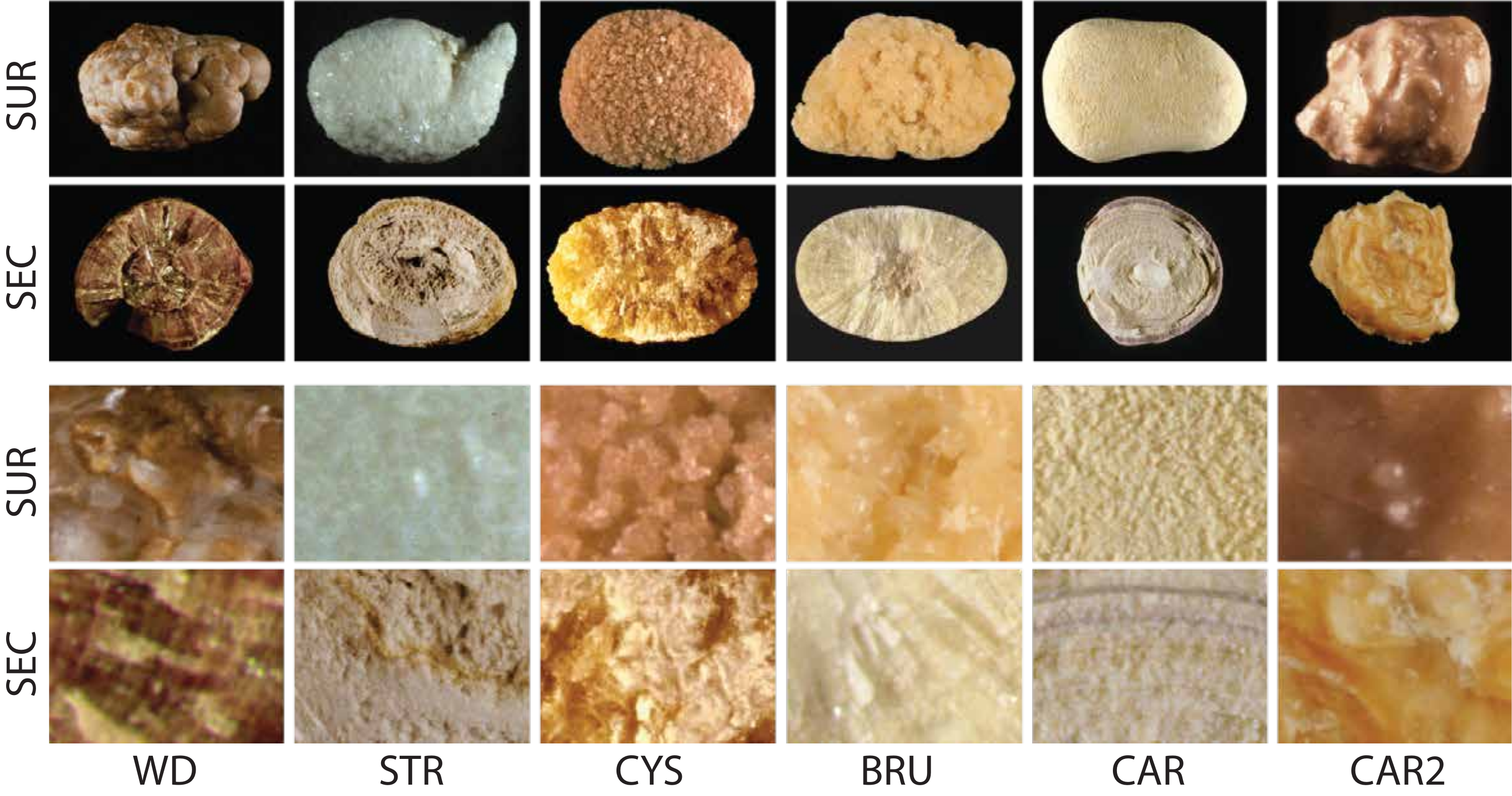}}
    \hspace{5mm}
    \subfloat[Dataset B: Endoscopic images (ex-vivo)]{
\label{fig:datasetb}\includegraphics[width=0.45\textwidth]{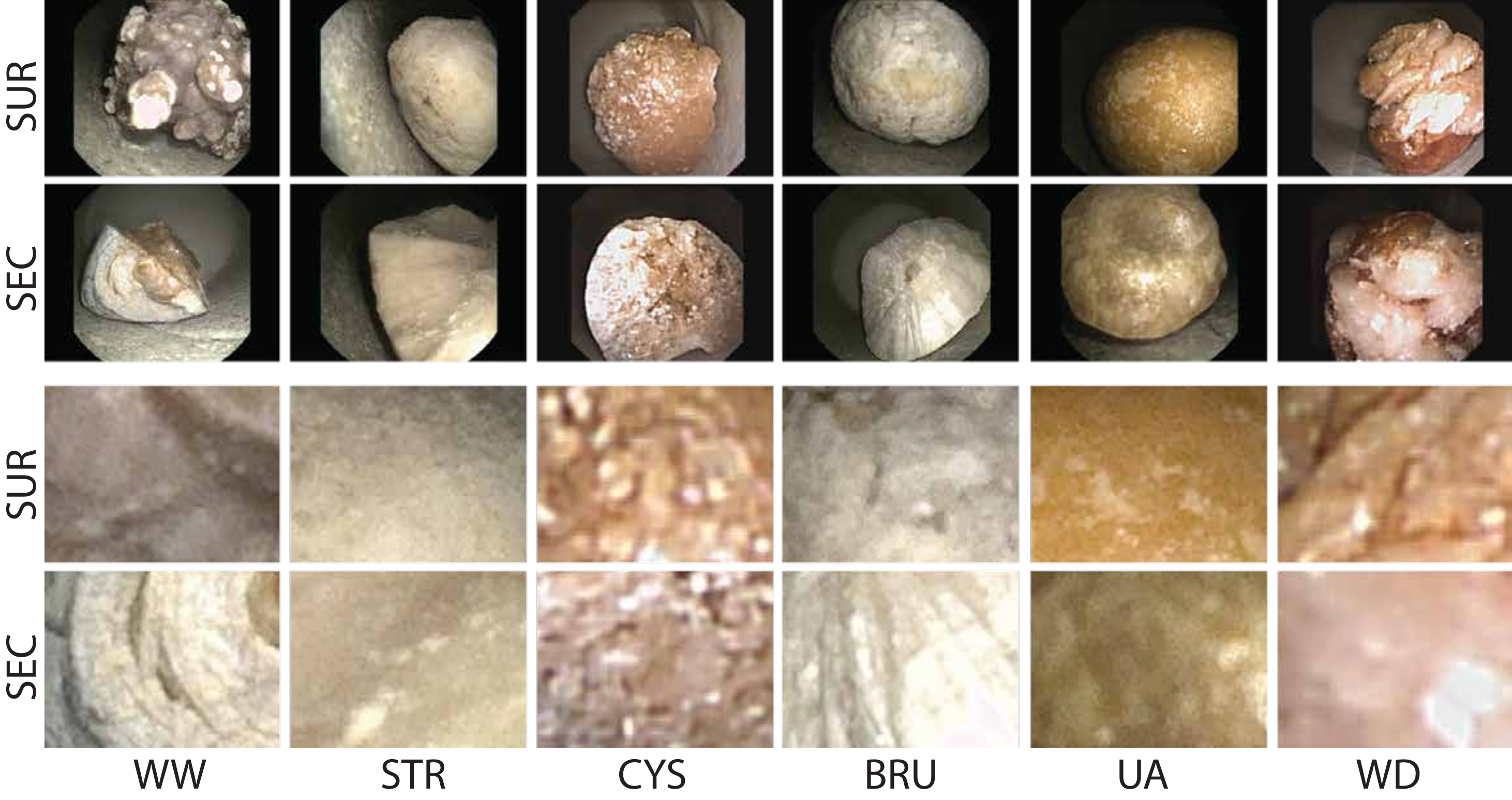}}

    \caption{Examples of ex-vivo kidney stone images acquired with (a) a CCD camera and  (b) an endoscope. SEC and SUR stand for section and surface views, respectively.} 
    \label{fig:dataset}
    \end{figure*}

%

\begin{table*}[]
\centering
\caption{Description of the two ex-vivo datasets.}
\label{tab:dataset}
\resizebox{\textwidth}{!}{%
\begin{tabular}{@{}ccccccccccc@{}}
\cmidrule(r){1-5} \cmidrule(l){7-11}
\multicolumn{5}{c}{Dataset A. M. DAUDON et al. \cite{corrales2021classification}}                            &  & \multicolumn{5}{c}{Dataset B. J. EL-BEZE et al. \cite{el2022evaluation}}                           \\ \cmidrule(r){1-5} \cmidrule(l){7-11} 
Subtype         & Main component (Key)         & Surface & Section & Total &  & Subtype         & Main component (Key)         & Surface & Section & Total \\ \cmidrule(r){1-5} \cmidrule(l){7-11} 
Ia              & Whewellite (WW)              & 50      & 74      & 124   &  & Ia              & Whewellite (WW)              & 62      & 25      & 87    \\
IVa1            & Carbapatite (CAR)            & 18      & 18      & 36    &  & IIa             & Weddelite (WD)               & 13      & 12      & 25    \\
IVa2            & Carbapatite (CAR2)           & 36      & 18      & 54    &  & IIIa            & Uric Acid (UA)           & 58      & 50      & 108   \\
IVc             & Struvite (STR)               & 25      & 19      & 44    &  & IVc             & Struvite (STR)               & 43      & 24      & 67    \\
IVd             & Brushite (BRU)               & 43      & 17      & 60    &  & IVd             & Brushite (BRU)               & 23      & 4       & 27    \\
Va              & Cystine (CYS)                & 37      & 11      & 48    &  & Va              & Cystine (CYS)                & 47      & 48      & 95    \\ \cmidrule(r){1-5} \cmidrule(l){7-11} 
\multicolumn{2}{c}{Number of images dataset A} & 209     & 157     & 366   &  & \multicolumn{2}{c}{Number of images dataset B} & 246     & 163     & 409   \\ \cmidrule(r){1-5} \cmidrule(l){7-11} 
\end{tabular}
}
\end{table*}
    
This paper has two contributions: i) it proposes a novel DL-model for fusing information included in endoscopic images of the two views (surface and section) of a kidney stone fragment with the aim to increase the discrimination performance and, ii) it shows how a multi-branch model can be trained using a two-step transfer learning (TL) approach in order to improve the model generalization capabilities.

This paper is organized as follows.
Section \ref{sec:sota}  reviews the literature on automated  ESR and introduces the key concepts used in this work, namely multi-view fusion and two-step TL.
Section \ref{sec:material-methods} describes the construction of the dataset, details the two-step TL setup, and presents the pre-training stage of the multi-view model.
Section \ref{sec:result-discussion} compares the results obtained with the proposed model in several configurations, with that of other models given in previous works.
Finally, section \ref{sec:conclusion} discusses future research directions.

\section{State-of-the-art}
\label{sec:sota}

Different DL approaches for an automated classification of kidney stones demonstrated encouraging results \cite{
lopez2021assessing,
ochoa2022vivo}. 
However, DL-models require large data amounts to yield accurate results  \cite{
mendez2022generalization, mehra2018breast}.
In ureteroscopy, it is difficult to collect such large datasets.  
A solution to this issue lies in methods such as TL and fine-tuning from other distributions (ImageNet) as a weight initialization technique  \cite{
lopez2022boosting}. 
Such techniques also enable to avoid training from scratch. 
However, for an automated endoscopic stone recognition (aESR), these initialization techniques are not useful, since the distribution of ImageNet and endoscopic (ureteroscopic) images are substantially different. Thus, customized TL methods that initialize useful weights closer to the target domain are required.

Furthermore, most models performing aESR were trained on surface or section images taken separately \cite{estrade2022towards, estrade2017should, mendez2022generalization}. However, the visual inspection in MCA (by biologists) and ESR (by urologists) is based on both views by jointly exploiting information from fragment surfaces and sections \cite{corrales2021classification, daudon2004clinical, daudon2016comprehensive}. So far, the DL-models in the literature did not use together surface and section information to improve the classification efficiency.
Multi-View (MV) classification is exploited in this contribution to combine the features observed in the two fragment-type views. 

The aim of this paper is to show that an MV-model outperforms models without an elaborated fusion strategy.  
MV is performed by fusing features (of shallow models) or feature maps (for DL-models) determined for various images with the aim to learn more complete representations and to obtain more effective classifiers \cite{villalvazo2022improved,su2015multi}. Contrary to a MV-approach, previous works for aESR were based on a DL-model, trained three times (only with section data, only for surface data, and for surface and section data gathered in the same class).
This contribution leverages recent advances in DL-based models that combine information from multiple viewpoints and improve the results using domain adaptation techniques.

\section{Materials and Methods}
\label{sec:material-methods}

\subsection{Datasets}
\label{sec:datasets}

Two kidney stone datasets were used in our experiments \cite{corrales2021classification, el2022evaluation}. 
According to the dataset, the images were acquired either with standard CCD cameras or with an ureteroscope (i.e., an endoscope). These datasets are described below.

\textbf{Dataset A,} \cite{corrales2021classification}.  This ex-vivo dataset of 366 CCD camera images (see, Figure \ref{fig:dataseta}, left column Table \ref{tab:dataset}) is split in 209 surface and 157 section images, and contains six different stone types sorted by sub-types denoted by WW (Whewellite, sub-type Ia), CAR (Carbapatite, IVa), CAR2 (Carbapatite, IVa2), STR (Struvite, IVc), BRU (Brushite, IVd), and CYS (Cystine, Va). 
The stone fragment images were acquired with a digital camera under controlled lighting conditions and  with a uniform background. 
The dimensions of the images in dataset A are 2848$\times$4288 pixels.

\textbf{Dataset B,} \cite{el2022evaluation}. The endoscopic dataset  consists of 409 ima\-ges (see Figure \ref{fig:datasetb}, right column Table \ref{tab:dataset}).
This dataset includes 246 surface and 163 section images. 
Dataset B involves the same classes as dataset A, except that the Carbapatite fragments (sub-types IVa1, and IVa2) are replaced by the Weddelite (sub-type IIa) and Uric Acid (IIIa) classes. 
The images of dataset B were captured with an endoscope by placing the kidney stone fragments in an environment simulating in a quite realistic way in-vivo conditions (for more details, see \cite{el2022evaluation}). The dimensions of the images in dataset B are 576$\times$768 pixels.

Automatic kidney stone classification is usually not performed on full images due to the limited size of the datasets. 
Therefore, as in previous works \cite{lopez2022boosting}, patches of 256$\times$256 pixels  were extracted from the original images to increase the size of the training dataset (for more details, see \cite{lopez2021assessing}).
However, it should also be mentioned whether the disadvantages have been taken into account, some of which are (i) loss of context since by cropping small regions of an image, contextual and spatial information may be lost, (ii) when patching an image, certain features may be present in multiple patches. On the other hand, one of the advantages is that by using patches it is possible to train machine learning models (which are complicated with few samples) and ensure an increase in the number of samples, and balance between classes.

A total of 12,000 patches were generated for each dataset which is organized into six classes as follows: For dataset A (WW, STR, CYS, BRU, CAR, CAR2) and dataset B (WW, WD,UA, STR, BRU, CYS). 
%
For each data set, $80\%$ of the patches (9600 patches) are used for the training and validation steps, while the remaining $20\%$ of the patches (2400 patches) act as test data (200 patch-images for each class). Patches of the same image contribute either only to the training/validation data or solely to the test data.
The patches were also ``whitened" using the mean $m_{i}$ and standard deviation $\sigma_{i}$ of the color values $I_{i}$ in each channel \cite{lopez2021assessing}.

\subsection{Proposed approach}

\begin{figure*}[!h]
  \begin{center}
    \includegraphics[width=1\textwidth]{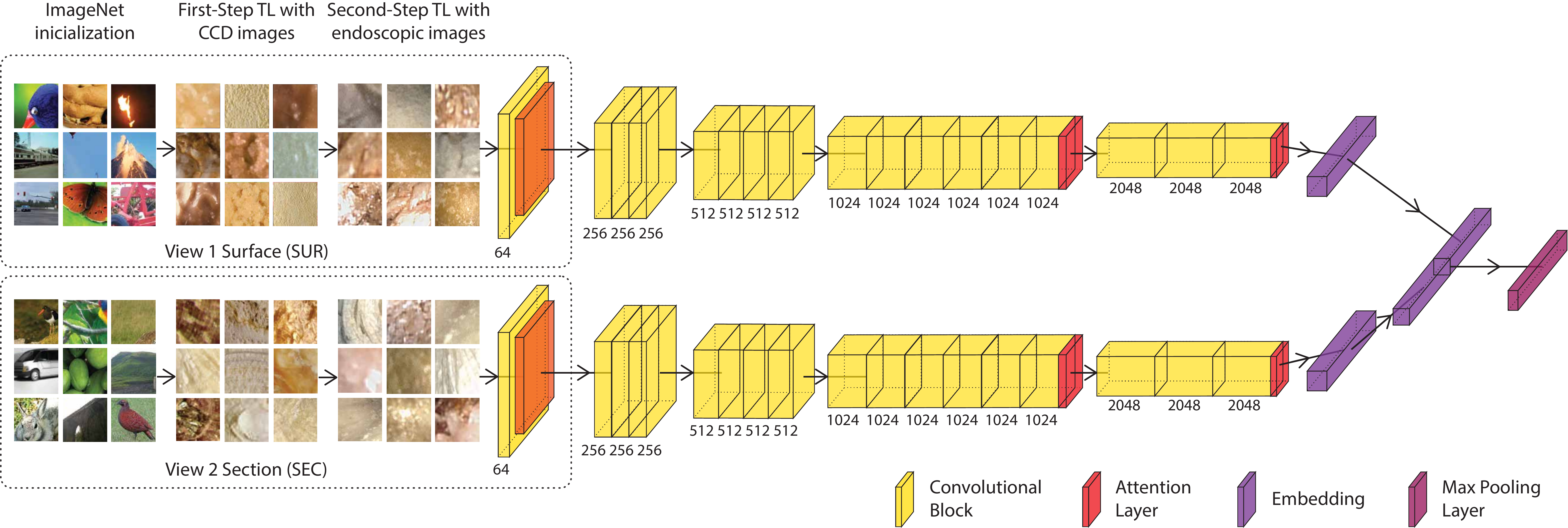}
  \end{center}

  \caption{Proposed multiview-fusion model assisted by two-step transfer learning for aESR.}
  \label{fig:model}
\end{figure*}

Several approaches \cite{lopez2021assessing}  have demonstrated the ability of DL-based models to recognize in single views (SUR or SEC) different types of kidney stones with high performance. However, in most cases, they have been trained by fine-tuning with a totally different distribution than kidney stones, or worse, they have been trained from scratch with the endoscopic images for individual views. 
On the other hand, although in the work \cite{lopez2021assessing}, it was observed that combining features (color and texture) from both views of an endoscopic kidney stone image (surface and section) produces more useful vectors for classification using shallow machine learning methods, or training surface and section patches together in a DL-based model produce more discriminative features compared to models trained with surface or section images, so far no elaborated technique was exploited to combine the surface and section information.

Usually, to exploit the information from SUR and SEC images, the patches of the two views of a fragment are simply seen as instances of the same class. Although such methods fuse both views information and more data available for the training, the way in which image features are extracted and combined is far from being optimal, as it does not emulate how the visual inspection of MCA/ESR is performed. To make matters worse, mixing the features in this way does not always improve the classification results. As can be observed in the MIX column of Table \ref{tab:comparison} (values marked by the * symbol), in some cases fusing features from SUR and SEC patches does not produce better feature maps, as this information combination is not optimal and hinders the model performance \cite{lopez2022boosting}.

In order to exploit the best features of both views, the proposed DL-model (see Figure \ref{fig:model}) combines the information in a systematic way using a fusion strategy based on a multi-view scheme, introducing attention mechanisms to further filter out unnecessary features maps of our CNN-model. Moreover, instead of training from the scratch the individual branches, we assist the model training with a two-step TL approach as a method of initializing weights from a similar distribution (CCD-camera images) to the endoscopic images.

\begin{figure} [h!]
     \centering
     \includegraphics[width=0.48\textwidth]{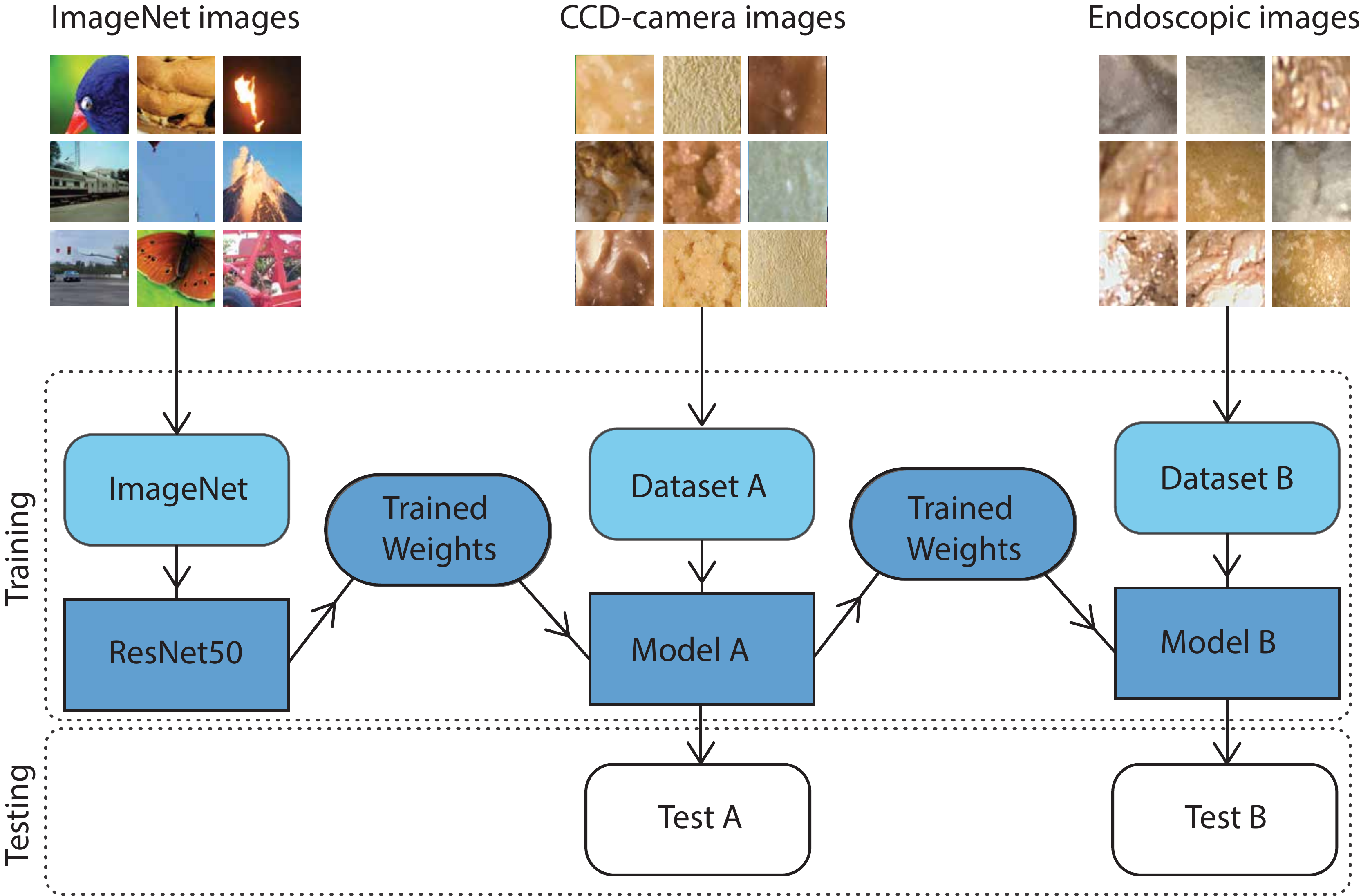}

   \caption{Two-step Transfer Learning. Model A was initialized with the weights of a ResNet50 network pre-trained with ImageNet, and fine-tuned with Dataset A. Next, Model B starts with the weights learned from Model A and is finally fine-tuned with Dataset B. The objective is to pre-train dataset B with weights similar to its distribution (kidney stone images).  }
     \label{fig:two-step}
     \end{figure}

\subsection{Two-step Transfer Learning} 
\label{sec:transfer}

There are different ways to acquire knowledge for a DL model (e.g. training from scratch). However, in applications where data scarcity is a constraint, techniques such as Two-step Transfer Learning (Figure \ref{fig:two-step}) are useful as a pre-training or initialization method for a  specific domain (dataset) \cite{lopez2022boosting}. 

Two-step Transfer Learning learns weights from different distributions approaching the final domain/target (endoscopic dataset). First, during the HeTL (HeTL stands for heterogeneous TL), the pre-training is performed with a general domain (ImageNet). 
The model weights are updated during a HoTL (homogeneous TL) using a domain whose data distribution is the closest to that of the target (domain adaptation process). In the kidney stone application, the pre-training  on ImageNet improves the generalization capabilities of the DL-model and the CCD camera images of ex-vivo fragments are used as a first fine-tuning. This fine-tuning is finalized using a part of the target dataset B (fragment images acquired with endoscopes), the remaining patches of dataset B being used for the validation and testing steps.
%
More specifically, during the HeTL-step, the large ImageNet dataset is used to transfer knowledge into a ResNet50 network which is fine-tuned by the smaller kidney stone image set acquired under controlled acquisition conditions (dataset A) as shown on the left part of Figure \ref{fig:model}.
%
Then, fine-tuning is achieved for each branch (i.e. individual model for each view) during the HoTL-step. This final tuning exploits dataset B which is composed of endoscopic images close to dataset A, but with higher variability in terms of image contrast, noise, and resolution, emulating thus the illumination and scene conditions actually encountered in ureteroscopy when patient data are acquired with an endoscope. The second-step TL is performed for each of the views (SUR/SEC) by obtaining two independent models trained with dataset B of endoscopic images for their respective views (for more details, see \cite{lopez2022boosting}). As described below, a MV-model, assisted by the second TL-step, is used to combine the SUR and SEC views into a mixed model (MIX).

\begin{table*}[]
\centering
\caption{Comparison of the performance of various kidney stone identification methods. The value of the accuracy over all classes was determined with dataset B for all methods.}
\label{tab:two-step}
\resizebox{\textwidth}{!}{%
\begin{tabular}{@{}cccccccl@{}}
\toprule
View & Dataset & TL Step & Accuracy    & Precision   & Recall      & F1-Score    & Training details                            \\ \midrule
     & B       & --      & 0.702$\pm$0.012 & 0.718$\pm$0.010 & 0.702$\pm$0.012 & 0.701$\pm$0.008 & Trained with endoscopic images              \\
SUR  & A       & 1       & 0.649$\pm$0.050 & 0.655$\pm$0.039 & 0.649$\pm$0.050 & 0.642$\pm$0.046 & Fine-tuning with ImageNet weights           \\
     & \textbf{B}       & \textbf{2}       & \textbf{0.832$\pm$0.012} & \textbf{0.845$\pm$0.012} & \textbf{0.832$\pm$0.012} & \textbf{0.829$\pm$0.012} & Fine-tuning with microscopic images weights \\ \midrule
     & B       & --      & 0.738$\pm$0.022 & 0.772$\pm$0.015 & 0.738$\pm$0.022 & 0.722$\pm$0.023 & Trained with endoscopic images              \\
SEC  & A       & 1       & 0.824$\pm$0.022 & 0.834$\pm$0.020 & 0.824$\pm$0.022 & 0.820$\pm$0.023 & Fine-tuning with ImageNet weights           \\
     & \textbf{B}       & \textbf{2}       & \textbf{0.904$\pm$0.048} & \textbf{0.915$\pm$0.037} & \textbf{0.904$\pm$0.048} & \textbf{0.903$\pm$0.050} & Fine-tuning with microscopic images weights \\ \midrule
     & B       & --      & 0.760$\pm$0.024 & 0.773$\pm$0.029 & 0.760$\pm$0.024 & 0.752$\pm$0.024 & Fine-tuning with ImageNet weights           \\ 
MIX  & A       & 1       & 0.800$\pm$0.013 & 0.809$\pm$0.013 & 0.800$\pm$0.013 & 0.797$\pm$0.013 & Fine-tuning with ImageNet weights           \\
     & \textbf{B}       & \textbf{2}       & \textbf{0.856$\pm$0.001} & \textbf{0.868$\pm$0.002} & \textbf{0.856$\pm$0.001} & \textbf{0.854$\pm$0.001} & Fine-tuning with microscopic images weights \\ \bottomrule
\end{tabular}%
}
\end{table*}

\begin{table*}[]
\centering
\caption{Comparison of different fusion strategies (concatenation and pooling) applying attention in different blocks (last and second last). }
\label{tab:fusion}
\begin{tabular}{@{}cccccl@{}}
\toprule
Fusion method & Accuracy   & Precision  & Recall     & F1-score   & Attention block            \\ \midrule
Concatenation & 91.00$\pm$3.03 & 92.00$\pm$3.16 & 91.00$\pm$3.03 & 91.00$\pm$3.03 & Without attention          \\
Pooling       & 90.40$\pm$3.29 & 90.80$\pm$3.11 & 90.40$\pm$3.29 & 90.40$\pm$3.29 & Without attention          \\
Pooling       & 88.33$\pm$6.43 & 88,33$\pm$5.51 & 88.00$\pm$6.08 & 88.00$\pm$6.08 & Last block                 \\
\textbf{Pooling}       & \textbf{91.25$\pm$0.50} & \textbf{91.75$\pm$0.50} & \textbf{91.25$\pm$0.50} & \textbf{91.25$\pm$0.50} & Last and second last block \\ \bottomrule
\end{tabular}
\end{table*}

\subsection{Multi-view model}
\label{sec:multiview}

Once the two SUR and SEC models are trained through the previous two-step TL, the feature extraction layers of this single-view network are frozen to ensure that each branch of the multi-view model extracts the same features and that any variation in performance depends on the non-frozen layers (merge and full connection layers).
These frozen layers are connected to a fusion layer, which is responsible for mixing the information of the two views. 

 In this work, the two late-fusion methods proposed for kidney stone classification  \cite{villalvazo2022improved} were exploited (concatenation and max-pooling). 
The first fusion method (concatenation) concatenates the feature vectors obtained from each view and merges the resulting representation through a fully connected layer. 
On the other hand, in the second fusion method (max-pooling), feature vectors are stacked and max-pooling is applied to them. 
Three configurations were used to implement max-pooling. The first corresponds to a model without attention mechanisms. The second consists of one layer of attention (last block). Finally, the third consists of two-layer of attention (the last, and second last block, arranged as shown in Figure \ref{fig:model}).
%
Lastly, the output of the late-fusion layer is connected to the remaining part of the MV-model, which merely consists of the classifier. The full proposed model is shown in Figure \ref{fig:model}.

\section{Results and Discussion}
\label{sec:result-discussion}

\subsection{Two-step Transfer Learning results}
\label{sec:results-twostep}

\begin{figure*} [] 
    \centering
    \subfloat[SUR ]{\label{fig:umap-sur}\includegraphics[width=0.32\linewidth]{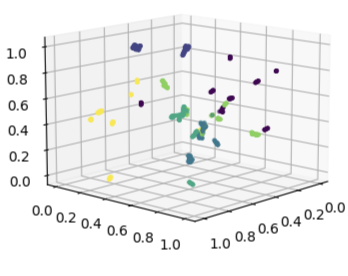}}
    \hspace{10mm}
    \subfloat[SEC]{
\label{fig:umap-sec}\includegraphics[width=0.32\linewidth]{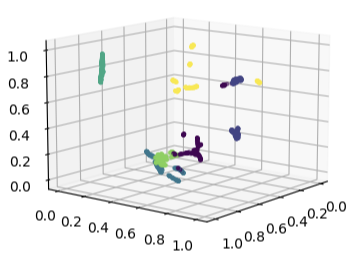}}

    \subfloat[MIX]{
\label{fig:umap-mix}\includegraphics[width=0.5\linewidth]{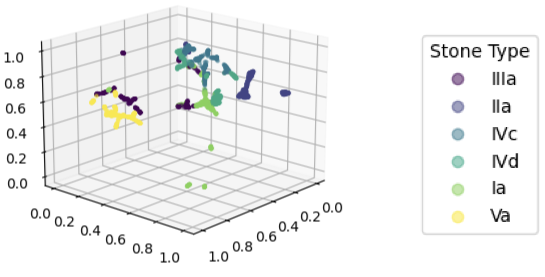}}

    \caption{UMAP \cite{mcinnes2018umap} dimensionality feature reduction (a) SUR,  (b) SEC, and (c) MIX patch sets.} 
    \label{fig:umap}
    \end{figure*}

Three experiments (SUR model, SEC model, MIX model) were carried out to assess the performance of the two-step TL approach applied to the patch data described in Section \ref{sec:datasets}. 
In the first and second experiments, the two-step TL approach described in Section \ref{sec:transfer} was used to predict kidney stone types in endoscopic images for individual SUR and SEC views. 
Then, in the third experiment, the SUR and SEC datasets are mixed into a single dataset denoted as MIX. The model is trained by taking into account the patches of both views during training. The results obtained are shown in Table \ref{tab:two-step}.

In order to describe the results, in this work we will use the Accuracy metric, since the testing set is balanced and all the metrics (Precision, Recall and F1-Score) follow the same trend (as shown in Table \ref{tab:two-step}).

\textbf{SUR model.} The Two-step Transfer Learning (fine-tuning with microscopic images weights) results for the SUR view patch dataset is 83.2$\pm$01.2 (in $\%$ measured with accuracy).  The performance of the Two-Step based model outperforms training from scratch with endoscopic images (70.2$\pm$01.2), and also greatly improves on the performance obtained by doing a single step of Transfer Learning with ImageNet (64.9$\pm$05.00\%). Although the latter result would be expected to be superior to training from scratch (with endoscopic images), we assume that the ImageNet weights for the SUR view are not useful for this task. Which gives us a slight hint that weight initialization is not always useful. 

\textbf{SEC model.} The results obtained in the SEC view (90.4$\pm$04.8) are promising using two-step Transfer Learning, and suggest better performance than the SUR view (83.2$\pm$01.2), probably due to the extraction of more discriminant features. In comparison, training a SEC model from scratch with endoscopic images (73.8$\pm$02.2), the performance improved significantly. On the other hand, pre-training with ImageNet (82.4$\pm$02.2) shows an increase in accuracy quite useful for this task. 

\textbf{MIX model.} In order to measure the performance of a model that considers both views during training, MIX model was evaluated. An overall yield of 85.6$\pm$0.10 was obtained by applying two-step TL to the MIX assembly. The results are good, with respect to training from scratch with endoscopic images (76.0$\pm$02.4) or fine-tuning with ImageNet (80.0$\pm$01.3). However, it would be expected that combining the information during training would yield better features than individual views (SUR 83.2$\pm$01.2 and SEC 90.4$\pm$04.8). These results for MIX suggest that combining information during training does not always generate good results. To deal with this problem, it is proposed to use the models trained on SUR and SEC datasets implemented with two-step Transfer Learning were combined using the MV-model described in Section \ref{sec:multiview}. 


\subsection{Multiview-fusion model}
\label{sec:results-multiview}

The individual SUR and SEC models trained with two-step Transfer Learning are used as two individual branches for the fusion scheme. In this work, two fusion methods were implemented: max-pooling and concatenation. For max-pooling, three variants were implemented (without attention, with attention in the last block, and with attention in the last two blocks). The results of these experiments are gathered in Table \ref{tab:fusion} and discussed below.

\textbf{Concatenation.} For a simple fusion technique that does not require attention mechanisms, promising results were obtained. For this technique, a performance of 91.00$\pm$3.03 was obtained in the inference of surface or section patches. With respect to the MIX model of section (\ref{sec:results-twostep}) implemented with two-step Transfer Learning , it can be observed that a significant improvement of up to 6\% accuracy.

\textbf{Pooling without attention.} The performance obtained with max-pooling without attention (90.40$\pm$3.29, max-pooling slightly below concatenation) as well as concatenation (91.00$\pm$3.03) maintains a similar accuracy. This suggests that both techniques of fusing information in an organized manner and assisted by two-step Transfer Learning, are a viable alternative for the classification of renal calculi in endoscopic images, outperforming the MIX model by a good margin (up to 6\%). 

\textbf{Pooling with attention (last block).} Under the premise of extracting better features from the individual SUR and SEC models, an attention mechanism was applied to the last block of the fusion scheme. The results for pooling with attention in the last block  (88.33$\pm$6.43), remained slightly below concatenation and max-pooling, and are still superior to MIX. 

\textbf{Pooling with attention (last and second last block).} In order to improve the performance of the max-pooling results, attention was applied to the last two blocks of the model (as shown in Figure \ref{fig:model}).  In addition, the results show an improvement over the other experiments with max-pooling and concatenation. The max-pooling with attention (last and second last block) approach is the model selected for comparison with the state of the art that will be described below.

\subsection{Comparison with the state-of-the-art}

In order to compare the results obtained in this work for SEC, SUR and MIX models, the works \cite{martinez2020towards, estrade2022towards, black2020deep, lopez2022boosting} were reimplemented and trained with the datasets described in Sec. \ref{sec:datasets}. The evaluation of the models was performed with the same testset. 

Regarding the SUR view, the performance obtained in this contribution using two-step Transfer Learning exceeds the performance obtained by related works. Qualitatively, we can observe in Figure \ref{fig:umap-sur}, that the features extracted and plotted by UMAP remain dispersed among elements of the same classes. However, the qualitative performance is good, surpassing the results of the state of the art which follows the same trend.

On the other hand, in the SEC view, the model presented in this work shows better results than those described in previous works, and maintains a superior performance. In Figure \ref{fig:umap-sec}, it can be observed that the features extracted by the SEC view are more discriminative with respect to the SUR view. Also, it can be seen that elements of the same class are grouped correctly, and that the extra-class distance is large, supporting the quantitative results of Table \ref{tab:two-step}.

Finally, the fusion scheme has been shown to be efficient  combining both views (SUR/SEC) in a "mixed" model. In addition the proposed model can maintain the performance, contrary to the models of state-of-the-art that do not have an organized way of combining information (marked by the * symbol).  The latter shows that combining the SUR and SEC information of stones in a single class leads to a performance decrease. Qualitatively (Figure \ref{fig:umap-mix}), the MIX view presents compact clusters for all classes; however, the inter-class distance is desired to increase. Applying attention techniques at deeper levels could improve these features, or implement them to the concatenation model.

\begin{table}[t!]
\centering
\caption{Comparison of the performance of various aESR DL-methods. The classification accuracy (in percentage) overall classes was determined with test dataset B for all methods.}

\label{tab:comparison}
\resizebox{\columnwidth}{!}{%
\begin{tabular}{@{}cccc@{}}
\toprule
Method & SUR & SEC & MIX \\ \midrule  
Martinez, et al. \cite{martinez2020towards} & 56.2$\pm$23.3 & 46.6$\pm$12.8 & *52.7$\pm$18.9 \\ 
Estrade, et al. \cite{estrade2022towards}  & 73.7$\pm$17.9 & 78.8$\pm$10.6 & *70.1$\pm$22.3 \\ 
Black, et al. \cite{black2020deep} & 73.5$\pm$19.0 & 76.2$\pm$18.5 & \textcolor{white}{*}80.1$\pm$13.8 \\ 
Lopez-Tiro, et al. \cite{lopez2021assessing} &81.0$\pm$03.0  &88.0$\pm$02.3  &*85.0$\pm$03.0 \\ 
\textbf{This contribution} & \textbf{83.2$\pm$01.2} & \textbf{90.4$\pm$04.8} & \textbf{\textcolor{white}{*}91.2$\pm$0.50} \\ \bottomrule 
\end{tabular}
}
\end{table}

\section{Conclusion and future work}\label{sec:conclusion}
\label{discussion_future_work}

This contribution shows that, by mixing information from two views, it is possible to train more accurate models to identify kidney stones acquired with endoscopes. Thus, AI technology can be an interesting solution for assisting urologists. However, these contributions used a very limited dataset in terms of class number and patch samples. The learning approaches on few samples must be improved to cope with the small amount of training data, and especially to increase the class separability when more kidney stone types have to be identified.  

\section*{Acknowledgements}
The authors wish to acknowledge the Mexican Council for Science and Technology (CONACYT) for the support in terms of postgraduate scholarships in this project, and the Data Science Hub at Tecnologico de Monterrey for their support on this project. 
This work has been supported by Azure Sponsorship credits granted by Microsoft's AI for Good Research Lab through the AI for Health program. The project was also supported by the French-Mexican ANUIES CONAHCYT Ecos Nord grant 322537.

{\small
\bibliographystyle{ieee_fullname}
\bibliography{egbib}

\begin{thebibliography}{10}\itemsep=-1pt

\bibitem{black2020deep}
Kristian~M Black, Hei Law, Ali Aldoukhi, Jia Deng, and Khurshid~R Ghani.
\newblock Deep learning computer vision algorithm for detecting kidney stone
  composition.
\newblock {\em BJU international}, 125(6):920--924, 2020.

\bibitem{cloutier2015kidney}
Jonathan Cloutier, Luca Villa, Olivier Traxer, and Michel Daudon.
\newblock Kidney stone analysis:“give me your stone, i will tell you who you
  are!”.
\newblock {\em World journal of urology}, 33(2):157--169, 2015.

\bibitem{corrales2021classification}
Mariela Corrales, Steeve Doizi, Yazeed Barghouthy, Olivier Traxer, and Michel
  Daudon.
\newblock Classification of stones according to michel daudon: a narrative
  review.
\newblock {\em European Urology Focus}, 7(1):13--21, 2021.

\bibitem{daudon2016comprehensive}
Michel Daudon, Arnaud Dessombz, Vincent Frochot, Emmanuel Letavernier,
  Jean-Philippe Haymann, Paul Jungers, and Dominique Bazin.
\newblock Comprehensive morpho-constitutional analysis of urinary stones
  improves etiological diagnosis and therapeutic strategy of nephrolithiasis.
\newblock {\em Comptes Rendus Chimie}, 19(11-12):1470--1491, 2016.

\bibitem{daudon2004clinical}
Michel Daudon and Paul Jungers.
\newblock Clinical value of crystalluria and quantitative morphoconstitutional
  analysis of urinary calculi.
\newblock {\em Nephron Physiology}, 98(2):p31--p36, 2004.

\bibitem{daudon2012stone}
Michel Daudon and Paul Jungers.
\newblock Stone composition and morphology: a window on etiology.
\newblock In {\em Urolithiasis}, pages 113--140. Springer, 2012.

\bibitem{daudon2018recurrence}
Michel Daudon, Paul Jungers, Dominique Bazin, and James~C Williams.
\newblock Recurrence rates of urinary calculi according to stone composition
  and morphology.
\newblock {\em Urolithiasis}, 46:459--470, 2018.

\bibitem{el2022evaluation}
Jonathan El~Beze, Charles Mazeaud, Christian Daul, Gilberto Ochoa-Ruiz, Michel
  Daudon, Pascal Eschw{\`e}ge, and Jacques Hubert.
\newblock Evaluation and understanding of automated urinary stone recognition
  methods.
\newblock {\em BJU international}, 2022.

\bibitem{estrade2022towards}
Vincent Estrade, Michel Daudon, Emmanuel Richard, Jean-christophe Bernhard,
  Franck Bladou, Gr{\'e}goire Robert, and Baudouin Denis~de Senneville.
\newblock Towards automatic recognition of pure and mixed stones using
  intra-operative endoscopic digital images.
\newblock {\em BJU international}, 129(2):234--242, 2022.

\bibitem{estrade2017should}
V Estrade, M Daudon, O Traxer, and P Meria.
\newblock Why should urologist recognize urinary stone and how? the basis of
  endoscopic recognition.
\newblock {\em PROGRES EN UROLOGIE}, 27(2):F26--F35, 2017.

\bibitem{estrade2017pourquoi}
V Estrade, M Daudon, O Traxer, P M{\'e}ria, et~al.
\newblock Pourquoi l’urologue doit savoir reconna{\^\i}tre un calcul et
  comment faire? les bases de la reconnaissance endoscopique.
\newblock {\em Progr{\`e}s en Urologie-FMC}, 27(2):F26--F35, 2017.

\bibitem{friedlander2015diet}
Justin~I Friedlander, Jodi~A Antonelli, and Margaret~S Pearle.
\newblock Diet: from food to stone.
\newblock {\em World journal of urology}, 33(2):179--185, 2015.

\bibitem{hall2009nephrolithiasis}
Phillip~M Hall.
\newblock Nephrolithiasis: treatment, causes, and prevention.
\newblock {\em Cleveland Clinic journal of medicine}, 76(10):583--591, 2009.

\bibitem{kartha2013impact}
Ganesh Kartha, Juan~C Calle, Giovanni~Scala Marchini, and Manoj Monga.
\newblock Impact of stone disease: chronic kidney disease and quality of life.
\newblock {\em Urologic Clinics}, 40(1):135--147, 2013.

\bibitem{kasidas2004renal}
GP Kasidas, CT Samuell, and TB Weir.
\newblock Renal stone analysis: why and how?
\newblock {\em Annals of clinical biochemistry}, 41(2):91--97, 2004.

\bibitem{khan2018fourier}
Aysha~Habib Khan, Sheharbano Imran, Jamsheer Talati, and Lena Jafri.
\newblock Fourier transform infrared spectroscopy for analysis of kidney
  stones.
\newblock {\em Investigative and Clinical Urology}, 59(1):32--37, 2018.

\bibitem{lopez2021assessing}
Francisco Lopez, Andres Varelo, Oscar Hinojosa, Mauricio Mendez, Dinh-Hoan
  Trinh, Yonathan ElBeze, Jacques Hubert, Vincent Estrade, Miguel Gonzalez,
  Gilberto Ochoa, et~al.
\newblock Assessing deep learning methods for the identification of kidney
  stones in endoscopic images.
\newblock In {\em 2021 43rd Annual International Conference of the IEEE
  Engineering in Medicine \& Biology Society (EMBC)}, pages 2778--2781. IEEE,
  2021.

\bibitem{lopez2022boosting}
Francisco Lopez-Tiro, Juan~Pablo Betancur-Rengifo, Arturo Ruiz-Sanchez, Ivan
  Reyes-Amezcua, Jonathan El-Beze, Jacques Hubert, Michel Daudon, Gilberto
  Ochoa-Ruiz, and Christian Daul.
\newblock Boosting kidney stone identification in endoscopic images using
  two-step transfer learning.
\newblock {\em arXiv preprint arXiv:2210.13654}, 2022.

\bibitem{martinez2020towards}
Adriana Mart{\'\i}nez, Dinh-Hoan Trinh, Jonathan El~Beze, Jacques Hubert,
  Pascal Eschwege, Vincent Estrade, Lina Aguilar, Christian Daul, and Gilberto
  Ochoa.
\newblock Towards an automated classification method for ureteroscopic kidney
  stone images using ensemble learning.
\newblock In {\em 2020 42nd Annual International Conference of the IEEE
  Engineering in Medicine \& Biology Society (EMBC)}, pages 1936--1939. IEEE,
  2020.

\bibitem{mcinnes2018umap}
Leland McInnes, John Healy, and James Melville.
\newblock Umap: Uniform manifold approximation and projection for dimension
  reduction.
\newblock {\em arXiv preprint arXiv:1802.03426}, 2018.

\bibitem{mehra2018breast}
Rajesh Mehra et~al.
\newblock Breast cancer histology images classification: Training from scratch
  or transfer learning?
\newblock {\em ICT Express}, 4(4):247--254, 2018.

\bibitem{mendez2022generalization}
Mauricio Mendez-Ruiz, Francisco Lopez-Tiro, Daniel Flores-Araiza, Jonathan
  El-Beze, Gilberto Ochoa-Ruiz, Miguel Gonzalez-Mendoza, Jacques Hubert, Andres
  Mendez-Vazquez, and Christian Daul.
\newblock On the generalization capabilities of fsl methods through domain
  adaptation: a case study in endoscopic kidney stone image classification.
\newblock In {\em Mexican International Conference on Artificial Intelligence},
  pages 249--263. Springer, 2022.

\bibitem{ochoa2022vivo}
Gilberto Ochoa-Ruiz, Vincent Estrade, Francisco Lopez, Daniel Flores-Araiza,
  Jonathan~El Beze, Dinh-Hoan Trinh, Miguel Gonzalez-Mendoza, Pascal
  Eschw{\`e}ge, Jacques Hubert, and Christian Daul.
\newblock On the in vivo recognition of kidney stones using machine learning.
\newblock {\em arXiv preprint arXiv:2201.08865}, 2022.

\bibitem{scales2012prevalence}
Charles~D Scales~Jr, Alexandria~C Smith, Janet~M Hanley, Christopher~S Saigal,
  Urologic~Diseases in America~Project, et~al.
\newblock Prevalence of kidney stones in the united states.
\newblock {\em European urology}, 62(1):160--165, 2012.

\bibitem{silva2010chemical}
Silvia Fernandes Ribeiro~da Silva, Djamile Cordeiro~de Matos, S{\^o}nia
  Leite~da Silva, Elizabeth De~Francesco Daher, Henry de~Holanda Campos, and
  Carlos Antonio Bruno~da Silva.
\newblock Chemical and morphological analysis of kidney stones: a double-blind
  comparative study.
\newblock {\em Acta Cirurgica Brasileira}, 25:444--448, 2010.

\bibitem{su2015multi}
Hang Su, Subhransu Maji, Evangelos Kalogerakis, and Erik Learned-Miller.
\newblock Multi-view convolutional neural networks for 3d shape recognition.
\newblock In {\em Proceedings of the IEEE international conference on computer
  vision}, pages 945--953, 2015.

\bibitem{viljoen2019renal}
Adie Viljoen, Rabia Chaudhry, and John Bycroft.
\newblock Renal stones.
\newblock {\em Annals of clinical biochemistry}, 56(1):15--27, 2019.

\bibitem{villalvazo2022improved}
Elias Villalvazo-Avila, Francisco Lopez-Tiro, Jonathan El-Beze, Jacques Hubert,
  Miguel Gonzalez-Mendoza, Gilberto Ochoa-Ruiz, and Christian Daul.
\newblock Improved kidney stone recognition through attention and multi-view
  feature fusion strategies.
\newblock {\em arXiv preprint arXiv:2211.02967}, 2022.

\end{thebibliography}

}

\end{document}